# Strategies and Influence of Social Bots in a 2017 German state election – A case study on Twitter

**Florian Brachten**
University of Duisburg-Essen
Duisburg, Germany
Email: florian.brachten@uni-due.de

**Stefan Stieglitz**
University of Duisburg-Essen
Duisburg, Germany
Email: stefan.stieglitz@uni-due.de

**Lennart Hofeditz**
University of Duisburg-Essen
Duisburg, Germany
Email: lennart.hofeditz@stud.uni-due.de

**Katharina Kloppenborg**
University of Duisburg-Essen
Duisburg, Germany
Email: katharina.kloppenborg@stud.uni-due.de

**Annette Reimann**
University of Duisburg-Essen
Duisburg, Germany
Email: annette.reimann@stud.uni-due.de

## Abstract

As social media has permeated large parts of the population it simultaneously has become a way to reach many people e.g. with political messages. One way to efficiently reach those people is the application of automated computer programs that aim to simulate human behaviour - so called social bots. These bots are thought to be able to potentially influence users' opinion about a topic. To gain insight in the use of these bots in the run-up to the German Bundestag elections, we collected a dataset from Twitter consisting of tweets regarding a German state election in May 2017. The strategies and influence of social bots were analysed based on relevant features and network visualization. 61 social bots were identified. Possibly due to the concentration on German language as well as the elections regionality, identified bots showed no signs of collective political strategies and low to none influence. Implications are discussed.

## 1   Introduction

Social media have gained importance in political communication over the last years. People as well as political actors use social media such as Twitter to debate political topics or to conduct political online campaigns (Yang et al. 2016). Twitters retweet system combined with its public nature strongly adds to the diffusion of information (Stieglitz and Dang-Xuan 2012). However, social media like Twitter also attract people who aim to abuse their functionalities and apply their potential as an efficient way to spread messages to a large audience with little effort (Rinke 2016). One potential danger in this context is that users could attempt to manipulate public opinion or to disrupt political communication. Within social networks such as Twitter, an effective tool for accomplishing this feat is the use of so called social bots (Woolley 2016).

Social bots are automated social media accounts designed to mimic human behavior (Abokhodair et al. 2015; Ferrara et al. 2016; Freitas et al. 2015). Through the simulation of human behaviour they are at first glance not easily recognizable as artificial accounts (Ferrara et al. 2016). This in turn could lead to human users misjudging the importance of the messages spread by such accounts eventually leading to being influenced in favour of the messages at display. The accounts differ on their level of sophistication with low-level-accounts, merely aggregating information from websites and using it to produce simple messages, e.g. on Twitter. A more sophisticated social bot on the other hand can be conversational and aim at passing as a human (Abokhodair et al. 2015).

The application of such accounts has been observed in several political contexts such as the Brexit debate in 2016 or the US presidential election in 2016 where social bots were responsible for roughly one-fifth of the conversation on Twitter (Howard and Kollanyi 2016). They potentially influenced users' opinion about the election as one candidate seemed to have more support than the other (Bessi and Ferrara 2016; Kollanyi et al. 2016). Kindled by observations of the use of these accounts in important votes, a debate considering the use of such accounts in the state election in 2017 in the most populous German state of North-Rhine Westphalia (NRW) has been a topic in the media and politics of the country (Rinke 2016). Driven by the ongoing debate about potential dangers of social bots and by the statement of the right wing populist party *Alternative für Deutschland* (*Alternative for Germany - AfD*) to potentially use social bots, all other major parties officially refrained from using social bots during their campaigns ("Das sagen die NRW-Parteien zu Social Bots" 2016). Accordingly, social bots in support of the right-winged *AfD* have been identified on Facebook by the popular media (Bender and Oppong 2017).

Besides the detection of social bots themselves, another important part in research is the detection and identification of different strategies social bots utilize while the attempt to take part in a discussion (Abokhodair et al. 2015). This is especially true for political context where social bots are thought to manipulate users' perceptions of political actors or parties (Bessi and Ferrara 2016; Woolley 2016). Furthermore, as social bots are a rather new phenomenon, the literature on social bot usage in German politics or votes is sparse. Therefore, one goal of this paper is to deepen the understanding of social bot use in political contexts and more specifically in Germany on the example of the NRW state election in May 2017. To do so, social bots on Twitter were detected and analyzed concerning their strategies and the influence they exerted on human users. Twitter was chosen for comparability with previous findings, as the majority of research on the topic of social bots takes place on Twitter (Stieglitz et al. 2017).

Following this introduction section 2 will present state of the art social bot detection techniques and findings on strategies and influence of social bots which will conclude in the research questions. Section 3 presents the dataset and methods used. Findings are presented in section 4 and discussed in section 5.

## 2   Theoretical Background

Among the research on social bots one of the largest streams are papers with the focus on the use of social bots in politics. The application of these bots could be shown in several different political scenarios such as Venezuelan politics in general (Forelle et al. 2015), the Syrian civil war (Abokhodair et al. 2015), US-American mid-term elections (Mustafaraj and Metaxas 2010) or the conflict between Ukraine and Russia in 2015 (Hegelich and Janetzko 2016). Furthermore, in two of the most far reaching votes in recent years, namely the 2016 UK vote to leave the European Union as well as the 2016 US presidential election, the application of bots was also detected (Bessi and Ferrara 2016; Howard and Kollanyi 2016). Especially as the outcome of the latter two events was contrary to what the polls had indicated in the forerun, the debate in the aftermath in parts also considered social media and within those social bots to have played a role in the outcome. If follows, that with the election of a new parliament in Germany in the fall of 2017 some concerns are raised as to what extent social bots might be used on this occasion as well. However, research about the use of social bots in German politics is rare. In a data memo, Neudert et al. (2017) analyzed the use of social bots during the German Federal Presidency Election in February 2017. The overall impact of social bots was rather small as the authors found only 22 automated accounts with a total of about 6,000 tweets out of more than 36,500 accounts with about 120,000 tweets. However, as the German Federal President is not elected directly by the people this election may have been of less interest for social bot use. As North-Rhine Westphalia is Germanys most populous federal state (with 17.9 million citizens) elections here are considered a pre-election test for overall Germany. Regarding the election of a new German parliament in fall of 2017, this paper aims at expanding the findings on social bot usage in German politics on the example of the 2017 German federal election in NRW. In order to compare the bots' behavior the first step is to assess the strategies applied by bots in earlier studies.

## 2.1 Social Bot Strategies in Politics

Woolley (2016) conducted a qualitative content analysis of English news articles on social bots. The author found that governments and other actors deployed social bots during elections, country-specific political conversations or crises and for preemptive online security purposes. The strategies that social bots used varied depending on country and political entity. One strategy of social bots was to disseminate pro-government or pro-candidate tweets to sway public opinion in their favour. An article of the New York Times described this strategy in a state prosecutors' allegations in South Korea, where over one million tweets were sent from social bots. Politicians also bought social bots to boost their follower lists. In the US presidential election in 2012 Mitt Romney acquired nearly 117,000 followers in about 24 hours through presumptive social bot activities (Woolley 2016). Another finding was that communication surrounding elections was targeted with so called *Twitter bombs*, i.e. the flooding of a hashtag used by opposing parties through a social botnet and thus barraging that hashtag. Those practices were labeled *smokescreening* or *misdirecting* by Abokhodair et al. (2015) who observed similar practices in the context of the Syrian civil war. Spam tactics were also used for attacking various other countries and commercial organizations. Another term often used in the context is the so called *astroturfing* - tweeting or retweeting a (political) opinion to suggest broad consensus around this opinion. In other words, astroturfing is a term for the simulation of an artificial grassroot movement.

Abokhodair et al. (2015) used a content analysis to identify the strategies used by social bots in the Syrian botnet. They searched for content that did not match the hashtags used to detect smoke screening or misdirecting. Chu et al. (2012) considered features like the URL frequency as an efficient detection feature for social bots by analyzing 500,000 accounts. They found that social bots tend to include URLs in tweets to redirect users to external content and argued that this is due to Twitter's character limit of 140 characters.

According to Zhang et al. (2013a), a good indicator for astroturfing is if there are many automated tweets for a topic or a hashtag. Bessi and Ferrara (2016) observed how social bots affect political discussion in the context of the 2016 US presidential election. Through analyzing the sentiments of tweets that used hashtags related to the presidential election the authors found social bots to produce more positive content in support of a candidate than humans do. This strategy resembles astroturfing, as it tries to generate the impression that a certain opinion is more strongly represented than another.

A study by Ratkiewicz et al. (2011) researched how Twitter can be exploited through astroturfing campaigns. To answer this, they analyzed Twitter data that they obtained from the astroturfing detection tool *Truthy*. They found that astroturfing campaigns had to be detected early, because after passing Twitter's spam detector they can do great harm in a political campaign.

To sum it up, social bots can be implemented with different objectives in mind. In political contexts, they are applied to boost the number of followers, influence public opinion or disrupt communication (Woolley 2016). These objectives are strongly linked to different strategies. To create a forged grass root support for a specific agenda social bots tweet or retweet certain opinions (astroturfing) (Abokhodair et al. 2015; Zhang, Zhang, et al. 2013). They disrupt communication by flooding hashtags with unrelated but similar content (smoke screening) or by guiding users' attention away from a topic (misdirection) (Abokhodair et al. 2015). However, these social bot strategies are still quite new and therefore there is further need for research on the identification of strategies in political context. As mentioned before, research lacks findings considering social bots in the political context in Germany and as such also considering the strategies applied. To close this gap, our first research question is:

RQ1: *Were there bot activities in the context of the state election 2017 of NRW on Twitter and which strategies were used by social bots?*

Besides the question of strategies social bots apply another important aspect to answer is to what extend these accounts can influences public opinion. As mentioned before, through so called astroturfing bot accounts are thought to artificially generate the impression that a certain opinion has stronger support than it really does. While this and other similar mechanisms may seem retraceable in in theory it is important to look at the actual findings in influence of social bots.

## 2.2　Influence of Social Bots

Social media can be seen as an important tool for advancing democratic discussions, as it is a very powerful communication medium (Bessi and Ferrara 2016) and facilitates information diffusion (Stieglitz and Dang-Xuan 2012). Thus, political messages on Twitter can potentially be spread quickly and reach a high number of users. Nevertheless, determining the influence of social media interaction on political decision making and behavior is hard to assess and current research offers mixed results. On one side, according to a study by Kushin and Yamamoto (2010), social media only played a minor role in affecting situational political involvement and political self-efficacy. On the other side, different studies show an influence of social media use on offline political participation and political expression (Gil de Zúñiga et al. 2014). But what happens when social bot strategies, like astroturfing and smoke screening, are used in order to manipulate the content in social media?

Considering social bots, the first step to determine their impact is to detect the reach of these accounts. Bessi and Ferrara (2016) explored the level of network embeddedness of social bots in the US presidential election in 2016 to ascertain the reach of the manipulations. They found that humans retweeted other humans and social bots equally as often indicating that social bots can be just as effective in spreading information as human's accounts are. The authors concluded that the presence of social bots has the potential to negatively affect democratic political discussion and thus to manipulate public opinion and endanger the integrity of elections. Human users retweeting social bots indicate difficulty for normal users to differentiate between social bot and human user which is in accordance with findings by Everett et al. (2016). The authors confronted Internet users with genuine comments generated by humans as well as automatically generated comments. In 50% of the cases the users could not differentiate whether a comment came from another user or was generated automatically. Similarly, Freitas et al. (2015) found an akin number of obtained followers between social bots that solely retweeted other users tweets compared with bots who retweeted and additionally produced automatic tweets. The authors concluded that this is due to users not being able to identify automatically generated tweets, possibly as a result of messages on Twitter being short and often informal or grammatically incoherent which makes the identification of automated content difficult for human users.

The effectiveness of social botnets for digital influence manipulation was investigated by Zhang et al. (2013b). The authors built their own social botnet for spam distribution on Twitter and tried to boost the influence of the network. To measure the digital influence, they used scores by three commercial digital-influence services (Klout, Kred, and Retweet Rank) that were based on how frequently accounts are retweeted and mentioned in a certain amount of days. In their experiments Zhang et al. (2013b) concluded that the three presented measurement methods of digital influence could easily be manipulated by social bots. The manipulation of the measurement by social bots is possible, because reach plays an important role and is often very high for social bot accounts.

These examples show that social bots in general are able to trick other users into acting as if the automated accounts were human. This might be the reason for human users retweeting social bots equally often as human users, thus spreading their content and in turn leading to people potentially being influenced by social bots (Bessi and Ferrara 2016).

In this paper, the digital influence is measured mainly through the reach of social bots. According to Bessi and Ferrara (2016) the digital influence in political contexts played an important role in the US presidential election 2016. In European countries like Germany research in this field is still sparse. Thus, the second research question is:

RQ2: *What influence did social bots have in the NRW state election 2017 campaigns on Twitter?*

The answers to these two questions will show 1. if social bots were used and 2. if so what influence they had in a German state election. Findings will add to the body of current research on the topic of social bots and will expand it with knowledge on the use in German politics and 2. broaden knowledge about the actual importance of social bots. The following section describes the research design used to answer these questions.

# 3 Research Design

To answer the research questions, we captured and analysed a dataset of Tweets concerning the NRW state election in May 2017.

## 3.1 Dataset

The dataset was collected from the 23rd of April to the 21st of May 2017 via the Twitter SEARCH API and covers a month of tweets about the NRW state election 2017 which took place on the 14th of May. The following general keywords marking tweets about the state election were used for the selection of tweets: *Landtagswahl* (the German expression for state election), *Landtagswahl NRW*, *#landtagswahl*, *#landtagswahlnrw*, *#ltwnrw* and *#ltwnrw17*. Furthermore, mentions of and tweets by the official accounts of parties in the state parliament of NRW at that time were tracked. Additionally, the account of the right wing populist party AfD was tracked as well, as there were indicators of social bots connected to the AfD (e.g. Bender and Oppong 2017). Thus *@gruenenrw*, *@CDUNRW_de*, *@fdp_nrw*, *@nrwspd*, *@PiratenNRW* and *@AlternativeNRW* were included in the dataset.

As the aforementioned keywords included the general term *#landtagswahl* and as a state election in another German federal state (namely Schleswig-Holstein) also took place in May 2017, the initial dataset included some tweets concerning the latter election which had to be removed. Hence, tweets including the keyword *ltwsh* or mentions of the official Schleswig-Holstein accounts of the local parties (*@Gruene_SH*, *@SH_CDU*, *@FDP_SH*, *@spdsh*, *@piratenparteish*, *@AfD_LV_SH*) were dismissed when they did not contain the additional keywords *NRW* or *Nordrhein*. This led to 6,166 tweets being excluded with a total of 182,995 tweets remaining. Out of the remaining tweets, 125,468 were retweets and 57,527 were original tweets. A total of 33,481 unique accounts were analyzed. Further descriptive statistics are presented in table 1.

| Keywords | No. of tweets | Keywords | No. of tweets |
|---|---|---|---|
| Landtagswahl | 18,595 | @gruenenrw | 13,370 |
| Landtagswahl NRW | 1,274 | @CDUNRW_de | 7,863 |
| #landtagswahl | 3,748 | @fdp_nrw | 10,045 |
| #landtagswahlnrw | 139 | @nrwspd | 6,541 |
| #ltwnrw | 76,673 | @PiratenNRW | 17,512 |
| #ltwnrw17 | 48,538 | @AlternativeNRW | 17,447 |

*Table 1. Descriptive statistics of the dataset. Number of tweets includes original tweets and retweets.*

## 3.2 Methods

For the detection of social bots, BotOrNot and Python version 3.6.1 in combination with R version 3.4.0 were used. BotOrNot is one of the first social bot detection interface for Twitter and employs supervised machine learning classification algorithms trained with 15,000 examples for social bot accounts and 16,000 examples for human accounts and more than 5 million tweets in total to assess whether an account is likely to be a bot - or not (Davis et al. 2016). In the past papers used this system to classify accounts in their gathered sample (Ferrara et al. 2016). The systems offers an open source API with a possible Python interface. As BotOrNot cannot detect accounts exhibiting a mixture of human and social bot characteristic the system should be combined with other approaches (Ferrara et al. 2016). Thus, BotOrBot was only the first part in the detection process described in the following section.

### 3.2.1 Detection of Social Bots in the Sample

The schematic description of the identification process is shown in figure 1. As a first step BotOrNot was utilized to detect social bots (I) as it has been tested on a large amount of data and generates reliable results for simple to medium sophisticated social bots (Davis et al. 2016). BotOrNot offers a public API endpoint which was accessed using Python. The detection program generates a score ranging from 0-1 with a higher score indicating stronger social bot-like behavior. For an account to be declared a social bot a score of at least 0.7 was decided on which is consistent with prior studies (e.g. Bessi & Ferrara 2016). A second check (II) was then conducted by analyzing six features that proved to be useful in detecting social bots (verified accounts, Tweet frequency, follower-to-followee-ratio, account language, account creation time, and profile picture – cf. Alarifi et al. 2016; Chu et al. 2012; Ferrara et al. 2016; Neudert et al. 2017). If an account displayed social bot-like behavior for a feature, the account scored 1 for this feature otherwise it was assigned 0. A summation of the feature scores results in a score ranging from 0 to 6 for each account. A higher score indicates a higher probability of the account being a social bot. Accounts with a score of at least 5 were classified as social bots.

To make sure that the classification was successful two manual checks were conducted by two coders each. The first manual check (III) concerned the accounts that were classified as social bots in step I and II. The goal here was to judge if the accounts were correctly classified as social bots. Additionally, 50 random accounts from the dataset were examined manually (IV) to judge if the automated detection may have missed social bot accounts. Accounts that were already examined in the first manual check were not included in the second manual check. Accounts that were classified as humans by both coders were given a 0, accounts that were classified as social bots by both coders were given a 1 and otherwise the account was given a 2. This results in a score of 0-2. Only those accounts that were consistently classified as social bots by the coders (score = 1) were declared as social bots – irrespective of their BotOrNot and feature analysis scores. Criteria for the manual checks were: Was the content of tweets original, intelligent and *human-like*, i.e. did the user report what he/she was doing or feeling (Chu et al. 2012)? Did tweets contain irony, sarcasm or jokes (Abokhodair et al. 2015; Ferrara et al. 2016)? Were there references to friends, family members, etc.? Was a tweet part of a conversation between friends (Abokhodair et al. 2015)? Was the profile information individualized (Freitas et al. 2015)? Did the account follow suspicious accounts? Did the account have suspicious followers (Chu et al. 2012; Ferrara et al. 2016)?

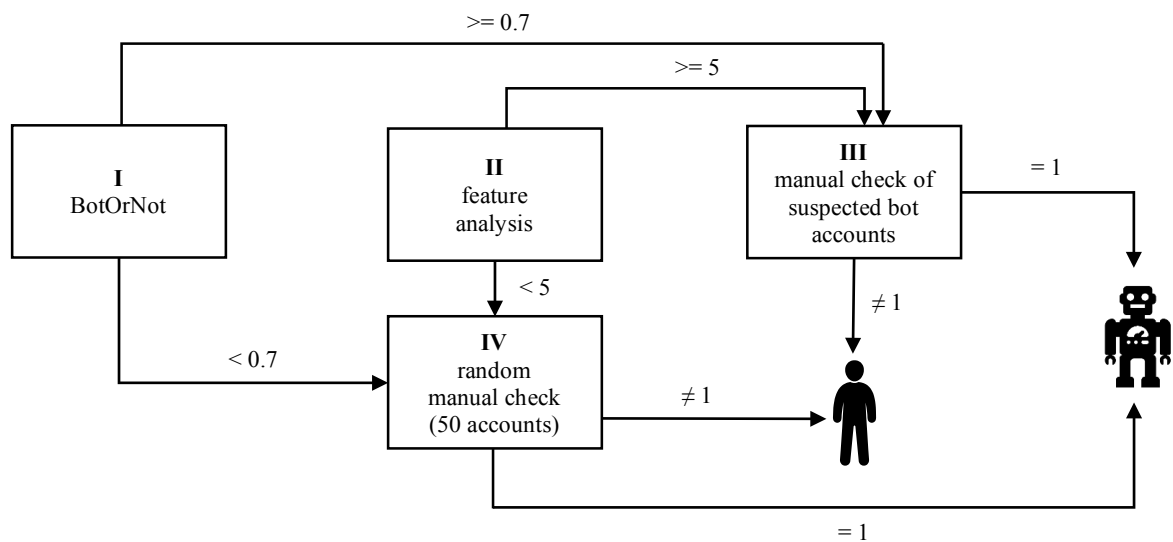

*Figure 1. Social bot detection procedure. Arabic numbers reflect the social bot detection scores described in section 3.2.1. The scores determined which path was taken. Roman numerals show the order in which the steps were conducted.*

Cohen's kappa adjusted by Byrt et al. (1993) was calculated to assess the consistency between the two coders. This adjusted version diminishes effects caused by prevalence, in this case unequal appearances of social bots and humans in the subsets. Both manual checks resulted in a satisfying inter-rater reliability ($k$ = 0.93 for the manual check of social bots and $k$ = 0.91 for the random check).

Through this procedure, a total of 61 social bot accounts could be identified which made up around 0.2% of all 33,481 accounts in the dataset.

### 3.2.2 Methods for Strategy Analysis

After detecting social bots their strategies were analyzed. First, a check for smoke screening and misdirecting was conducted. The two strategies were combined as they are not completely distinguishable from each other. For a tweet to be classified as smoke screening / misdirection the tweet content had to be unrelated to the hashtag used but still related to the topic of the state election or politics in general (Abokhodair et al. 2015). If the content was completely unrelated to the hashtag and had no connection to politic topics the tweet was classified as spam. The content of the tweets was checked manually by one coder. Additionally, the URL frequency of social bot accounts was calculated, i.e. the total number of URLs used by social bots was divided by the total number of tweets by social bots (Chu et al. 2012).

Furthermore, the tweets were checked for astroturfing, i.e. conveying support of a certain opinion or party. One characteristic of astroturfing is a high number of automated tweets concerning a certain topic (Zhang et al. 2013a) Therefore, the most used hashtags and keywords by social bots as well as by human users were determined to identify relevant topics in these groups. To analyze if topics were artificially pushed by social bots a ratio was calculated by dividing the number of times a popular hashtag was used by the number of tweets. This was done for human users and social bots individually to gain a metric that allowed comparison of hashtag frequency between humans and social bots. The same procedure was applied to popular keywords.

Automated tweets that could not be identified as smoke screening / misdirecting, spam or astroturfing were analyzed for similarities in order to uncover other strategies.

### 3.2.3 Methods for Influence Analysis

Finally, the influence of social bots was evaluated by assessing their reach. A retweet rank was calculated to uncover the most influential accounts. The rank was calculated sorting the accounts by the number of times their original tweets were retweeted by other users. To determine if human users retweeted social bots, it was distinguished whether those users were other social bots or human users.

## 4 Results

### 4.1 Behaviour of Detected Social Bots

The 61 identified social bot accounts tweeted 336 times in total with 199 messages being original tweets and 137 being retweets. In comparison, human users in the dataset tweeted 182,659 times during the data collection period. Of these, 121,664 were original tweets and 60,995 retweets. On average, social bots posted $M = 5.60$ ($SD = 14.00$, range = 1 - 96) tweets during the collection period, while human users *posted* $M = 5.47$ ($SD = 19.41$, range = 1 – 680). The values of the standard deviation and range indicate that the tweet frequency for social bots as well as for human users is skewed in that few users produced the most content. More precisely, the three most active social bot accounts tweeted 96, 40, and 35 times each in the dataset with *Rambonelli*, an AfD supporting social bot, being the most active. The most active human accounts tweeted 680, 609, and 553 times each and were accounts associated with the pirate party. The account *AlternativeNRW* ranks on fifth place while accounts of other parties were being less active.

### 4.2 Detected Strategies

No indication for the occurrence of smoke screening or misdirecting could be found as hashtags or URLs were either related to the content or completely unrelated. Those tweets containing URLs and hashtags unrelated to the topic were classified as spam. This strategy was discovered 24 times in the social bot dataset. However, the tweet frequency was too low to speak of an extensive strategy. URLs that were related to the content often led to news articles of various official and unofficial media sites. This behavior, together with the aforementioned spamming, is also reflected in a high URL frequency of 0.95 URLs per bot tweet.

The low overall occurrence of social bots in the dataset also led to the absence of astroturfing as this strategy constitutes of large numbers of tweets in favour of a certain topic. Said behavior could not be observed. The top five hashtags used by social bots were *#ltwnrw*, *#afd*, *#ltwnrw17*, *#nrw* and *#nachrichten*. The first four of these were also in the top five hashtags used by human accounts, but in a different order. On average humans shared the most used hashtag in 40% of their tweets. The same cannot be said for social bots. Here, the top hashtags were only used in 10% of the tweets each (tables 2, 3 and 4). The top keywords used by social bots were *Landtagswahl*, *NRW*, *Nordrhein-Westfalen*, *SPD* and *CDU* (the both biggest parties, competing for the lead in the vote). *Landtagswahl* was used in 42% of the tweets on average but concerns the state election in general rather than being connected to a certain party. As for *SPD* and *CDU* those keywords were used rather seldom. Additionally, humans used very similar top keywords (tables 5 and 6).

|  | Human accounts | Social Bots |
|---|---|---|
| Number of Tweets | 182,659 | 336 |
| Number of Hashtags | 396,796 | 597 |

Table 2: Tweet statistics for human and social bot accounts

| Top 5 Hashtags H | Count | Ratio |
|---|---|---|
| #ltwnrw | 71,428 | 0.39 |
| #ltwnrw17 | 46,858 | 0.26 |
| #nrw | 21,378 | 0.12 |
| #afd | 21,104 | 0.12 |
| #piraten | 7,815 | 0.04 |

Table 3: Top hashtags used in human tweets.

| Top 5 Hashtags SB | Count | Ratio |
|---|---|---|
| #ltwnrw | 59 | 0.11 |
| #afd | 53 | 0.10 |
| #ltwnrw17 | 50 | 0.09 |
| #nrw | 38 | 0.07 |
| #nachrichten | 18 | 0.03 |

Table 4: Top hashtags used in social bot tweets.

Nonetheless, some social bot accounts posted party-specific content. One example for this behavior was exhibited by the account *Rambonelli* which had a follower count of 1,081 a status count of 34,982 statuses (96 in this dataset) and a friend count of 1,024. The social bot tweeted or retweeted exclusively content that was related to topics of the AfD like "*RT @MdLFernandes: +++ Die neue Fahne der Arbeiter ist blau ! +++ Bei der Landtagswahl haben die meisten Arbeiter die #AfD gewÃ«hlt.... httpsâ€¦*"[sic!] (meaning 'The new colour of workers is blue! Most blue collar workers voted for the #AfD in the state election').

| Top 5 Keywords | Count | Ratio |
|---|---|---|
| NRW | 13,977 | 0.08 |
| Landtagswahl | 12,399 | 0.07 |
| Wählen | 8,660 | 0.05 |
| CDU | 5,631 | 0.03 |
| SPD | 5,514 | 0.03 |

Table 5: Top keywords used in human tweets.

| Top 5 Keywords | Count | Ratio |
|---|---|---|
| Landtagswahl | 142 | 0.42 |
| NRW | 94 | 0.28 |
| Nordrhein-Westfalen | 45 | 0.13 |
| SPD | 32 | 0.10 |
| CDU | 32 | 0.10 |

Table 6: Top keywords used in social bot tweets.

### 4.3 Detected Influence

For the detection of the influence of social bots in our dataset a retweet rank was calculated. From all detected social bot accounts only three produced original tweets were retweeted: Those were *afdwtal* on rank 904 with one original tweet and six retweets, all of them by human accounts, *pilgerpetrus* on rank 1,550 with one original tweet and three retweets, all of them by human accounts and *Immoprofil_JM* on rank 1,753, also with one original tweet, two retweets, all of them by social bot accounts. The most retweeted accounts in the whole dataset were *PiratenNRW*, *WDR* and *tagesschau* (the two latter both being news outlets) with 1,240, 1,172 and 1,070 retweets respectively.

## 5 Discussion

As the results show we only found very little social bot activity in our study (RQ1). Out of 33,481 unique accounts only 61 accounts were classified as social bots. Furthermore, social bot accounts on average posted only 5.6 tweets during the whole collection period. These results differ from research concerning the US presidential election where "political bot activity reached an all-time high" (Kollanyi et al. 2016, p4.) with 15% of the studied accounts being classified as social bots and the discovered social bots made up a significant part of the conversation (Bessi and Ferrara 2016). Concerning the first research question which strategies were used by social bots in the NRW state election 2017 campaigns on Twitter (RQ1) it can be stated that no collective strategies could be detected. While there were some social bot accounts that shared URLs to news articles the referenced websites were diverse and mostly concerned the state election in general instead of pushing news related to a specific party. Therefore, the underlying objective of these social bots does not seem concerned with manipulating communication about the NRW state election by shedding a positive or negative light on specific political parties or actors. In addition to the sharing of news articles indicators for spam were found for individual social bots. These social bots used hashtags concerning the state election and added content and / or URLs that were unrelated to the topic. As the content and URLs were completely unrelated to the state election and politics in general these tweets were not classified as smoke screening / misdirecting. Thus, social bots did not try to influence communication by hiding content associated with a certain hashtag behind unrelated political content. They also did not try to direct users' attention to other political topics as was the case for the Syrian botnet analyzed by Abokhodair et al. (2015). Both the sharing of news and spamming are reflected in the URL frequency which indicates that every social bot generated tweet on average contained an URL.

Another strategy, that was expected to be used by social bots during the NRW state election, was astroturfing. As indicated by the low hashtag and keyword ratios, social bots did not try to push a topic collectively. Additionally, more than half of the top 5 hashtags and keywords were related to the state election in general which does not indicate the creation of artificial support for a certain party. There were, however, individual accounts sending sophisticated propaganda which Woolley (2016) described as a characteristic of social bots in political contexts. In the NRW state election dataset just a few social bots used this strategy namely *Rambonelli* or *afdwtal*. Both accounts are in support of the AfD that was accused of using social bots and artificially increasing their follower number and reach (Bender and Oppong 2017; Reuter 2017). However, we did not find a network of social bots that used this strategy. One possible explanation for the weak activities is that the use of social bots is in a test phase and should not arouse much attention. Furthermore, the language as well as the regional nature of the state election could be a reason for the apparent low application of social bots. Most other studies that could show a higher degree of bot activity concentrated on large events in English-speaking countries (e.g. the Brexit vote or the US election). However, literature also presents examples for application in non-English speaking contexts (Abokhodair et al. 2015) as well as smaller scale political events (US mid-term elections - Mustafaraj and Metaxas 2010). Nevertheless, both of the latter examples either used English language or were in the context of larger political events (for example the Syrian civil war). This could mean that if neither of these criteria is at hand (as in the current example), the application of social bots may not be seen as an efficient tool as the costs for a potential influence could be too high. One possible way to test this assumption is to analyze the social media communication surrounding the election of a new parliament for whole Germany and then compare these findings to our current study. With the latter election being a larger political event, comparable to the vote for a new government in other countries, the usage of social bots could be seen as being more worthwhile than on the regional level.

In accordance with the findings on strategies used by social bots in the NRW state election, regarding the influence of social bots in the NRW state election 2017 campaigns (RQ2) it can be derived that social bots exerted almost no influence. Only three of the social bot accounts were retweeted at all and their highest retweet rank was 904 with six retweets for *afdwtal*. Nevertheless, these retweets as well as those from *pilgerpetrus* were all carried out by human accounts which means that there was some human-social bot interaction, but to a very small extent. *Immoprofil_JM* was not retweeted by human accounts but twice by other social bots. However, this small number can rather be evaluated as coincidence than evidence for a social botnet. This is supported by the social network analysis that was conducted based on Bessi and Ferrara (2016) who explored the level of network embeddedness to measure the reach of accounts. It turned out that in the current sample social bots were largely outside the main graph which means that they were hardly networked in this dataset. Most of the connections seemed to be outside the graph. In this respect, the results differ greatly from those of Bessi and Ferrara (2016), who found many social bots within their social graph of the US presidential elections 2016.

As with all research some limitations to our findings apply. First, one important part of every study on social bots is the identification of bots in a dataset. While we attempted to apply a method which was at the same time efficient to use as well as effective it is possible that among the accounts classified as social bots there are some human users and the other way around. Also, while we relied on an external tool which promises to use over 1,000 different variables in the process of identification, it is still possible that BotOrNot is not able to detect German bots as it may be with other languages. Here, further research is needed, if maybe bots are sophisticated in a way that they differ depending of the country they are targeted at. The other important aspect to consider is that we explicitly only concentrated on German accounts, not regarding English tweets that may have also occurred within our dataset and may have enlarged the body of social bots and thus their influence. However, a closer look into the data did not indicate that this was the case.

In summary, only few social bots were present in the Twitter communication about the NRW state election 2017. These social bots did not pursue collective strategies and were hardly networked with one another or with humans.

## 6 Conclusion

The present study investigated the use of social bots, their strategies and influence in political contexts on the example of the NRW state election 2017 on Twitter. Contrary to findings from different countries and elections this study could reveal only a small number of accounts showing enough signs of automation that they could be classified as social bots. The only strategies that could be identified were traditional spamming without a political concept and the sharing of diverse news articles concerning the state election. In general, social bots were only loosely connected to the network and to each other and no social botnets were identified. This is already indicative for a low reach of social bots in the dataset and is further supported by social bots being seldom retweeted. These results deviating from other research in political contexts could be due to the language used (German) as well as the regional nature of the state election. Future studies thus could a) validate these findings on different elections on a similar level and b) compare these findings to other political events in Germany of the same size as well as larger votes such as the elections to the Bundestag.

Consequently, according to this study there are almost no social bots, let alone political bots used in the NRW state election campaigns 2017 on Twitter. One reason for these findings the German language that makes it harder for international social bot developers to create social bot campaigns for political events in Germany.

Even though the results indicated a very small use of social bots with no collective strategies and low reach more research on social bot use in Germany is desirable. As argued before the NRW state election might simply not have been of enough importance for social bots to be deployed. One interesting use case is the upcoming German federal election 2017 as it is the biggest political election in Germany. At this event, the practice using social bots on Twitter or other social media like Facebook should be analyzed. Especially Facebook should be considered due to the popularity of it in Germany.

## 7 References

Abokhodair, N., Yoo, D., and McDonald, D. W. 2015. "Dissecting a Social Botnet," in *Proceedings of the 18th ACM Conference on Computer Supported Cooperative Work & Social Computing - CSCW '15*, pp. 839–851 (doi: 10.1145/2675133.2675208).

Alarifi, A., Alsaleh, M., and Al-Salman, A. 2016. "Twitter turing test: Identifying social machines," *Information Sciences*, (372), pp. 332–346 (doi: 10.1016/j.ins.2016.08.036).

Bender, J., and Oppong, M. 2017. "Frauke Petry und die Bots," *Frankfurter Allgemeine Zeitung Online* (available at http://www.faz.net/aktuell/politik/digitaler-wahlkampf-frauke-petry-und-die-bots-14863763.html; retrieved July 31, 2017).

Bessi, A., and Ferrara, E. 2016. "Social bots distort the 2016 U.S. Presedential election online discussion," *First Monday*, (21:11), pp. 1–14.

Byrt, T., Bishop, J., and Carlin, J. B. 1993. "Bias, prevalence and kappa," *Journal of Clinical Epidemiology*, (46:5), pp. 423–429 (doi: 10.1016/0895-4356(93)90018-V).

Chu, Z., Gianvecchio, S., Wang, H., and Jajodia, S. 2012. "Detecting automation of Twitter accounts: Are you a human, bot, or cyborg?," in *IEEE Transactions on Dependable and Secure Computing*, (Vol. 9), pp. 811–824 (doi: 10.1109/TDSC.2012.75).

"Das sagen die NRW-Parteien zu Social Bots.," 2016. *Westdeutscher Rundfunk* (available at http://www1.wdr.de/nachrichten/landespolitik/parteien-nrw-social-bots-100.html; retrieved July 31, 2017).

Davis, C. A., Varol, O., Ferrara, E., Flammini, A., and Menczer, F. 2016. "BotOrNot: A System to Evaluate Social Bots," *Proceedings of the 25th International Conference Companion on World Wide Web*, (: 1602.009), pp. 273–274 (doi: 10.1145/2818717).

Everett, R. M., Nurse, J. R. C., and Erola, A. 2016. "The anatomy of online deception: What makes automated text convincing?," in *Proceedings of the ACM Symposium on Applied Computing*, (Vol. 04–08–Apri), pp. 1115–1120 (doi: 10.1145/2851613.2851813).

Ferrara, E., Varol, O., Davis, C., Menczer, F., and Flammini, A. 2016. "The rise of social bots," *Communications of the ACM*, (59:7), pp. 96–104 (doi: 10.1145/2818717).

Forelle, M. C., Howard, P. N., Monroy-Hernandez, A., and Savage, S. 2015. "Political Bots and the Manipulation of Public Opinion in Venezuela," *SSRN Electronic Journal*, pp. 1–8 (doi:

10.2139/ssrn.2635800).

Freitas, C. A., Benevenuto, F., Ghosh, S., and Veloso, A. 2015. "Reverse engineering socialbot infiltration strategies in twitter," in *Proceedings of the 2015 IEEE/ACM International Conference on Advances in Social Networks Analysis and Mining*, pp. 25–32 (doi: 10.1145/2808797.2809292).

Gil de Zúñiga, H., Molyneux, L., and Zheng, P. 2014. "Social Media, Political Expression, and Political Participation: Panel Analysis of Lagged and Concurrent Relationships," *Journal of Communication*, (64:4), Wiley Subscription Services, Inc., pp. 612–634 (doi: 10.1111/jcom.12103).

Hegelich, S., and Janetzko, D. 2016. "Are social bots on Twitter political actors? Empirical evidence from a Ukrainian social botnet," in *Proceedings of the Tenth International Conference on Weblogs and Social Media (ICWSM-2016)*, Palo Alto, CA: The AAAI Press, pp. 579–582.

Howard, P. N., and Kollanyi, B. 2016. "Bots, #StrongerIn, and #Brexit: Computational Propaganda during the UK-EU Referendum," (available at http://arxiv.org/abs/1606.06356).

Kollanyi, B., Howard, P. N., and Woolley, S. C. 2016. "Bots and automation over twitter during the second U.S. presidential debate.," *Comprop data memo*, (2).

Kushin, M. J., and Yamamoto, M. 2010. "Did Social Media Really Matter? College Students' Use of Online Media and Political Decision Making in the 2008 Election," *Mass Communication and Society*, (13:5), pp. 608–630 (doi: 10.1080/15205436.2010.516863).

Mustafaraj, E., and Metaxas, P. 2010. "From obscurity to prominence in minutes: Political speech and real-time search," *WebSci10: Extending the Frontiers of Society On-Line*, p. 317 (available at http://repository.wellesley.edu/computersciencefaculty/9/).

Neudert, L.-M., Kollanyi, B., and Howard, P. N. 2017. "Junk News and Bots during the German Federal Presidency Election: What Were German Voters Sharing Over Twitter?," Oxford (available at http://comprop.oii.ox.ac.uk/wp-content/uploads/sites/89/2017/03/What-Were-German-Voters-Sharing-Over-Twitter-v6-1.pdf).

Ratkiewicz, J., Conover, M. D., Meiss, M., Gonc, B., Flammini, A., and Menczer, F. 2011. "Detecting and Tracking Political Abuse in Social Media," *Icwsm*, pp. 297–304 (doi: 10.1145/1963192.1963301).

Reuter, M. 2017. "Größter AfD-Twitter-Account ist ein Scheinriese," *Der Tagesspiegel* (available at http://www.tagesspiegel.de/medien/datenjournalismus/twitter-datenanalyse-groesster-afd-twitter-account-ist-ein-scheinriese/19691492.html; retrieved July 31, 2017).

Rinke, A. 2016. "Hintergrund - Sorge um gekaufte digitale Hilfe im Wahlkampf 2017," (available at http://de.reuters.com/article/deutschland-parteien-sozialemedien-idDEKCN12L0EJ; retrieved July 31, 2017).

Stieglitz, S., Brachten, F., Ross, B., and Jung, A.-K. 2017. "Do Social Bots Dream of Electric Sheep? A Categorisation of Social Media Bot Accounts," *Proceedings of the Australasian Conference on Information Systems*.

Stieglitz, S., and Dang-Xuan, L. 2012. "Political communication and influence through microblogging - An empirical analysis of sentiment in Twitter messages and retweet behavior," *Proceedings of the Annual Hawaii International Conference on System Sciences*, pp. 3500–3509 (doi: 10.1109/HICSS.2012.476).

Woolley, S. C. 2016. "Automating power: Social bot interference in global politics," *First Monday*, (21:4) (doi: 10.5210/fm.v21i4.6161).

Yang, X., Chen, B.-C., Maity, M., and Ferrara, E. 2016. "Social Politics: Agenda Setting and Political Communication on Social Media," in *Social Informatics: 8th International Conference, SocInfo 2016, Bellevue, WA, USA, November 11-14, 2016, Proceedings, Part I*, E. Spiro and Y.-Y. Ahn (eds.), Cham: Springer International Publishing, pp. 330–344 (doi: 10.1007/978-3-319-47880-7_20).

Zhang, J., Carpenter, D., and Ko, M. 2013. "Online astroturfing: A theoretical perspective," *19th Americas Conference on Information Systems, AMCIS 2013*, (4:1), pp. 2559–2565 (available at http://www.scopus.com/inward/record.url?eid=2-s2.0-84893308397&partnerID=40&md5=f6d9be2d64336eb6f42f53f56d4e253b).

Zhang, J., Zhang, R., Zhang, Y., and Yan, G. 2013. "On the impact of social botnets for spam distribution and digital-influence manipulation," in *2013 IEEE Conference on Communications and Network Security, CNS 2013*, pp. 46–54 (doi: 10.1109/CNS.2013.6682691).